\documentclass[pre,floats,superscriptaddress,showpacs,usenames]{revtex4}
\usepackage{amsmath}
\usepackage{graphicx}

\usepackage[dvipsnames]{color}

\newcommand{\be}{\begin{equation}} 
\newcommand{\ee}{\end{equation}}
\newcommand{\bea}{\begin{eqnarray}}   
\newcommand{\eea}{\end{eqnarray}}

\newcommand{\rr}{{\bf r}}
\newcommand{\nvec}{\boldsymbol{n}}

\newcommand{\F}{\boldsymbol{F}}
\newcommand{\colxi}{\boldsymbol{u}}
\newcommand{\Fu}{{\cal F}}
\newcommand{\Ha}{{\cal H}}

\newcommand{\UU}{{\cal U}}

\newcommand{\eeta}{\boldsymbol{\eta}}

\begin{document}
\date{\today}


\title{Effective potential method for active particles}

\author{Umberto Marini Bettolo Marconi\footnote[3]
{(umberto.marinibettolo@unicam.it)}
}

\address{ Scuola di Scienze e Tecnologie, 
Universit\`a di Camerino, Via Madonna delle Carceri, 62032 ,
Camerino, INFN Perugia, Italy}

\author{Matteo Paoluzzi}

\address{Department of Physics, Syracuse University, Syracuse NY 13244, USA}

\author{Claudio Maggi}

\address{  Dipartimento di Fisica, Universita Sapienza, Rome, Italy  }

\begin{abstract}

We investigate the steady state properties of an  active fluid modeled as an assembly of soft repulsive spheres subjected to Gaussian colored noise. Such a noise captures one of the salient aspects of active particles, namely the persistence of their motion and determines a variety of novel features with respect to familiar passive fluids. We show that within the so-called multidimensional unified colored noise approximation, recently introduced in the field of active matter, the model can be treated by methods similar to those employed in the study of standard molecular fluids. The system shows a tendency of the particles to aggregate even in the presence of purely repulsive forces because the combined action of colored noise and interactions enhances the the effective friction between nearby particles. We also discuss whether an effective two-body potential approach, which would allow to employ methods similar to those of density functional theory, is appropriate. The limits of such an approximation are discussed.
\end{abstract}

\maketitle

\section{Introduction}
Recently there has been an upsurge of interest towards the behaviour of the so-called  active fluids 
whose elementary constituents are either self-propelled due their ability to convert energy into motion, for instance by
chemical reactions, or receive the energy and impulse necessary to their motion when in contact with living matter, such as bacteria~\cite{bechinger2016,marchetti2013hydrodynamics,marchetti2015structure,poon2013clarkia,romanczuk2012active,
elgeti2014physics}. Examples of active matter systems include self-propelled colloids, swimming bacteria,  biological motors, swimming fish and flocking birds. The phenomenology of active fluids is quite different
from that characterizing out-of-equilibrium molecular fluids, often referred as passive fluids, and for this reason the field is so fascinating.  
The theoretical studies are based on  phenomenological models, constructed by using physical
insight, with the aim to reproduce the complex biological mechanisms or  chemical reactions 
determining the observed dynamics \cite{cates2012diffusive,bialke2014negative,fily2012athermal,stenhammar2013continuum}.
Among these models we mention the important work of Cates, Tailleur and coworkers  \cite{tailleur2008statistical,cates2013active} who elaborated the Run-and-Tumble (RnT) model of Berg \cite{berg2004coli} and Schnitzer \cite{schnitzer1993theory}.  Such a model is based on the observations that the trajectories of individual bacteria consist of relatively straight segments (runs) alternated by erratic motions which cause the successive pieces of trajectory to be in almost random relative directions (tumbles). The persistence length of the trajectory sets the crossover between a ballistic regime at short time scales and a diffusive regime at longer times.
Similarly chemically propelled synthetic Janus colloids have a persistent propulsion
direction which is gradually reoriented by Brownian fluctuations~\cite{zheng2013nongaussian} (active Brownian particles). Also an ensemble of  colloidal particles suspended in a "bath" of  such bacteria, is a particular realization of an active fluid~\cite{maggi2014generalized,koumakis2014delivery,koumakis2013targeted,maggi2011effective}.
Our description, at variance with  the Cates and Tailleur model and the active Brownian model, involves only translational degrees of freedom of the particles,
but not their orientations, and represents somehow a coarse-grained version of these models
as also discussed by Farage et al.~\cite{farage2015effective,marconi2015towards}.
In the present model in order to  capture the peculiar character of the  RnT motion
on a coarse-grained time scales one introduces a colored noise, that is noise with a finite memory, which represents the persistence of the motion of the bacteria.
  
The observed behaviour displays a relevant feature: the particles  display a spontaneous tendency to aggregate even in the absence of mutual attractive forces, as a result of the combined effect of colored noise and interactions. This is a dynamic mechanism leading to a decrease of the particle effective mobility when the  density increases.

Clearly this behaviour  cannot be described by standard equilibrium statistical mechanics, but it is possible to make progress in our understanding by applying kinetic methods and the theory of stochastic processes \cite{marconi2010dynamic} . 

This paper is organized as follows:  in Sec. \ref{Model} we present  the coarse-grained stochastic model 
describing an assembly of active particles, consisting of a set of coupled Langevin equations for the coordinates of the particles
subject to colored Gaussian noise. After switching from the Langevin description to  the corresponding Fokker-Planck equation 
we obtain  the stationary  joint probability distribution of $N$ particles  within the multidimensional unified colored noise approximation  (MUCNA) \cite{hanggi1995colored,cao1993effects,maggi2015multidimensional,marconi2015velocity}.
The resulting configurational distribution function can be written explicitly for a vast class of inter-particle potentials
and shows the presence of non-pairwise effective interactions due to the coupling between direct forces
and the colored noise. To reduce the complexity of the problem 
in section \ref{Effectivepotential} we discuss whether it is possible to further simplify the theoretical study by introducing a pairwise effective potential.  We analyze such an issue analytically by means of a system of just two particles in an external field and 
numerically using a one-dimensional system of $N$ soft repulsive spheres.
Our results show the limits of the effective two-body potential method. Finally in the last section we present our conclusions. Two appendices are included:  in appendix \ref{multipletimescale} we derive  the UCNA approximation by means of multiple time-scale analysis and   in  \ref{appendixdeterminant}  we obtain the  key approximation of the theory  necessary to  evaluate  the
functional determinant of the MUCNA in the case of a many-particle system.

 \section{Model system}
 \label{Model}  
  
In this section, we briefly describe  the salient assumptions employed to formulate the model adopted in the present work. First consider an assembly of  $N$ over-damped active Brownian particles at positions $\rr_i$, self-propelling with constant velocity $v_0$ along orientations $\nvec_i$, which change in time according to the stochastic law
\be
\dot{\nvec}_i =\sqrt{D_r} \eeta_i\times\nvec_i
\label{orientation}
\ee
where $\eeta_i(t)$ are Gaussian random processes distributed with zero mean,
time-correlations	
$\langle\eeta_i(t)\eeta_j(t')\rangle=2\boldsymbol{1}\delta_{ij}\delta(t-t')$, and
$D_r=1/\tau$  is a rotational diffusion coefficient. In addition
the particles experience  deterministic forces $\F_i =\!-\nabla_i \UU$,  generated by the
potential energy  $ \UU$.   The 
resulting governing equations are:
\be
\dot{\rr}_i = v_0\,\nvec_i  + \gamma^{-1}\F_i ,
\label{full_langevin}
\ee
where $\gamma$ is the friction coefficient.
  The resulting dynamics are persistent on short time scales, i.e.  the trajectories maintain their orientation for an average time $\tau$,
  and diffusive on larger  time scales. Hydrodynamic interactions are disregarded for the sake of simplicity together with inertial effects
  because particles are typically in a low-Reynolds-number regime \cite{purcell1977life}.

 Equations \eqref{orientation}  and
\eqref{full_langevin} involve the dynamics of both translational and rotational degrees of freedom  and are not a practical starting point  for developing a microscopic  theory. 
 Hence, it is   convenient  to  switch to a coarse-grained description stochastically equivalent to the original one on times 
 larger than $\tau$ . To this purpose, one   integrates out the angular coordinates
 as shown by Farage et al.\cite{farage2015effective}.
 According to this approximation, one introduces a colored stochastic noise term acting on the position coordinates 
and  replacing the stochastic rotational dynamics  \eqref{orientation}.
 The effective evolution equations are:
\begin{align}\label{effective_langevin}
\dot{\rr}_i(t) &= \frac{1}{\gamma}\F_i(t)  + \colxi_i(t), 
\end{align}
with
\be\label{exp_correlation}
\dot\colxi_i(t) =- \frac{1}{\tau}\colxi_i(t) + \frac{ D^{1/2}}{ \tau} \eeta_i(t)  
\ee 
where $\colxi_i(t)$ is an Ornstein-Uhlenbeck process  with zero mean, time-correlation function  given by:
\be
\label{exp_correlation2}
\langle\colxi_i(t)\,\colxi_j(t')\rangle =  \frac{D}{\tau} e^{-2 |t-t'|/\tau}\boldsymbol{1}\delta_{ij}.   
\ee
and whose diffusion coefficient $D$ is related to the original parameters by $D=v_0^2 \tau /6 $.
 In order to derive an equation involving only the positions of the particles we differentiate with respect to time eq.\eqref{effective_langevin}
 and with simple manipulations we get the following second order differential equation:
 \bea
\ddot x_i= \frac{1}{\gamma} \sum_k \frac{\partial F_i}{\partial x_k} \dot x_k 
-\frac{1}{\tau}  \left[ \dot x_i-\frac{F_i}{\gamma} \right]       +\frac{D^{1/2}}{\tau}\eta_i(t) 
\label{sistema}
\eea
 where 
for the sake of notational economy we indicated by $x_i$ the array $\{\rr_i\}$ and similarly the components of the force.  
By performing an adiabatic approximation (see appendix \ref{multipletimescale}) we neglect the terms $\ddot x_i$ and obtain 
the following set of Langevin equations for the particles coordinates:
\be
\dot x_i \simeq \sum_k \Gamma^{-1}_{ik}  \Bigl[ \frac{1}{\gamma}  F_k+
D^{1/2}  \eta_k(t)   \Bigl]
\label{stochasticxi}
\ee
with the non dimensional friction  matrix $\Gamma_{ik}$ defined as
\be
\Gamma_{ik}(x_1,\dots,x_N)=\delta_{ik}+\frac{\tau}{\gamma} \frac{\partial^2 \UU(x_1,\dots,x_N) }{\partial x_i \partial x_k}.
\label{gammamatrix}
\ee
Notice that,  within the approximation introduced in eq. \eqref{stochasticxi}, the effective random force corresponds to a multiplicative noise due to its dependence on the state of the system, $x_i(t)$, through  the prefactor  $\Gamma^{-1}_{ik}(x_1,\dots,x_N)$ in front of the noise term $\eta_k(t)$.

For the sake of concreteness 
 $\UU$ is the sum of one-body and two body contributions:
 \be
\UU(x_1,\dots,x_N) =\sum_i u(x_i)+ \sum_{i>j} w(x_i,x_j) .
\ee
The associated multidimensional  Smoluchowski equation for the
the configurational distribution function  associated with eq. \eqref{stochasticxi} 
can be written as (see ref. \cite{marconi2015towards}):
\bea
\frac{ \partial P_N(x_1,\dots,x_N;t)}{\partial t}=-\sum_{l}  \frac{\partial }{\partial x_l} \sum_k \Gamma^{-1}_{lk} \Bigl(\frac{1}{\gamma } F_k  P_N 
- D \sum_{j}      \frac{\partial }{\partial x_j}
 [ \Gamma^{-1}_{jk}   P_N ] \Bigr)
\label{FPE}
\eea
and shows that  the effective friction 
experienced by each particle also
depends on the coordinates of all other particles.  
In order to determine the stationary properties of the model
 we apply the following zero current condition in eq. \eqref{FPE} and get:
 \bea
 && -T_s \sum_\beta \sum_n \frac{\partial }{\partial r_{\beta n}}[  \Gamma^{-1}_{\alpha l,\beta n}(\rr_1,\dots,\rr_N) P_N(\rr_1,\dots,\rr_N)]
 \nonumber\\&&=    
   P_N(\rr_1,\dots,\rr_N) \Bigl(  \frac{\partial u(\rr_{\alpha l})}{\partial r_{\alpha l}} +  \sum_{k\neq l}   \frac{ \partial w(\rr_l-\rr_k) }{ \partial r_{\alpha l}}   
\Bigl) \, .
\label{FPEstationary}
\eea 
The resulting  stationary distribution can be written explicitly as
(see ref.  \cite{maggi2015multidimensional}):
  \be
   P_N(x_1,\dots,x_N)  = \frac{1}{Z_N} \, \exp \Bigl(- \frac{\Ha(x_1,\dots,x_N)}{T_s} \Bigr)
   \label{probabilitydistr}
    \ee
where we have defined the effective temperature $T_s=D\gamma$ and the effective configurational energy of the system $\Ha(x_1,\dots,x_N)$ related  to the bare potential energy
$\UU(x_1,\dots,x_N)$  by:
\be
\Ha(x_1,\dots,x_N) = \UU(x_1,\dots,x_N)  +\frac{\tau}{2 \gamma}
    \sum_k^N \Bigr(\frac{\partial \UU(x_1,\dots,x_N)}{\partial x_k} \Bigl)^2 
    - T_s  \ln  |\det   \Gamma_{ik}  | ,
    \label{hamilt1}
\ee
  where  $Z_N$ is a normalization constant
  \be
Z_N=\mathrm{Tr} \exp \left[- \frac{\Ha(\rr_1,\dots,\rr_N)}{T_s} \right] .
\ee
with 
$\mathrm{Tr} \equiv \int d\rr_1 ,\dots,d\rr_N$.
Formula \eqref{probabilitydistr}, gives within the unified colored noise approximation, a complete information about the
configurational state of a system of $N$ particles, but
it requires the evaluation of a $dN\times dN$ determinant stemming from the matrix $\Gamma_{ik}$, where $d$ is the dimensionality
of the system. One can  only get analytic results either by considering non-interacting systems with $d=1,..,3$ or
systems with few particles, where the computation of the determinant is possible. 
Thus, in spite of the fact that in principle  from the knowledge of $P_N$ is possible to determine all steady properties  of the system, including  the pair correlation
function of the model, $g(\rr_1,\rr_2)$, this task is not possible in practice. The same situation occurs in equilibrium statistical mechanics where from the knowledge
of the canonical Boltzmann distribution of an $N$ particle system we cannot in general exactly determine the n-particle distribution functions with $n<N$.  On the other hand,
it is possible to derive a structure similar to the Born-Bogolubov-Green-Yvon
(BBGY) hierarchy of equations  linking the n-th order distribution to those of higher order, but it requires
the specification of a closure relation. To this purpose,
we integrate eq. \eqref{FPEstationary}
over  $d(N-n)$ coordinates and obtain an equation for the marginalized probability distributions of $n$ particles, $P_N^{(n)}(\rr_1,\dots,\rr_n)$ in terms of higher order marginal distributions. When $n=1$ we find:
\bea
&&
  T_s \int \int d\rr_2\dots d\rr_N \sum_{\beta=1}^d \sum_{n=1}^N\frac{\partial }{\partial r_{\beta n}}[  \Gamma^{-1}_{\alpha 1,\beta n}(\rr_1,\dots,\rr_N) P_N(\rr_1,\dots,\rr_N)]=    
    \nonumber\\
   && 
 -  P^{(1)}_N(\rr_1)  \frac{\partial u(\rr_{ 1})}{\partial r_{\alpha 1}} 
    - (N-1) \int d\rr_2   P^{(2)}_N(\rr_1,\rr_2)   \frac{ \partial w(\rr_1-\rr_2) }{ \partial r_{\alpha 1}}   
\label{p2distrb}
\eea  
where Greek indexes stand for Cartesian components.

In the case of a large number of particles the exact matrix inversion necessary to use formula \eqref{p2distrb}
becomes prohibitive. However, we notice
that in the limit of small $(\tau/\gamma)$ and $N\to\infty$  the structure  of the matrix $\Gamma^{-1}_{\alpha 1,\beta n}$  
becomes much simpler as illustrated in appendix \ref{appendixdeterminant}  and can be approximated by 
$$
\Gamma^{-1}_{\alpha l,\beta n}(\rr_l) \approx \Bigl( \delta_{\alpha\beta}-\frac{\tau}{\gamma} u_{\alpha\beta}(\rr_l)-
\frac{\tau}{\gamma}  \sum_{k\neq l}  w_{\alpha\beta}(\rr_l-\rr_k) ) \Bigl)\delta_{ln},
$$ 
where $u_{\alpha\beta}\equiv \frac{\partial^2 u(\rr)}{\partial r_\alpha \partial r_\beta}$ and
$w_{\alpha\beta}\equiv \frac{\partial^2 w(\rr)}{\partial r_\alpha \partial r_\beta}$.
Substituting this approximation in eq. \eqref{p2distrb} we get:
\bea
&&
 T_s  \sum_\beta \frac{\partial }{\partial r_{\beta 1}}\Bigl[P^{(1)}_N(\rr_1)  \delta_{\alpha\beta} -    \frac{\tau}{\gamma}    P^{(1)}_N(\rr_1)
  u_{\alpha\beta}(\rr_1) - (N-1)  \frac{\tau}{\gamma} \int \sum_k d\rr_2  P^{(2)}_N(\rr_1,\rr_2) w_{\alpha\beta}(\rr_1-\rr_2) \Bigl]
  \nonumber\\
   && 
  =-  P^{(1)}_N(\rr_1)  \frac{\partial u(\rr_{ 1})}{\partial r_{\alpha 1}} 
   -(N-1) \int d\rr_2   P^{(2)}_N(\rr_1,\rr_2)   \frac{ \partial w(\rr_1-\rr_2) }{ \partial r_{\alpha 1}}   
\label{p2bgy}
\eea  
Such an equation, once a prescription for  $P^{(2)}_N(\rr_1,\rr_2)$ is specified, can be used to derive the density profile of a system of interacting particles under inhomogeneous conditions. Let us remark that eq. \eqref{p2bgy}
expresses the condition of mechanical equilibrium equivalent to the first member of the BBGY  hierarchy as discussed in ref.
\cite{marconi2015towards}.

\section{Effective potential}
\label{Effectivepotential}

Let us apply eq.\eqref{FPEstationary}
   to a system  a system comprising just two 
particles so that the equation becomes closed:
\bea
&&
  T_s \sum_{\beta=1}^d \sum_{n=1}^2 \frac{\partial }{\partial r_{\beta n}}[  \Gamma^{-1}_{\alpha 1,\beta n}(\rr_1,\rr_2) P^{(2)}_2(\rr_1,\rr_2)]=    
  - P^{(2)}_2(\rr_1,\rr_2) \Bigl(  \frac{\partial u(\rr_{ 1})}{\partial r_{\alpha 1}} +  \frac{ \partial w(\rr_1-\rr_2) }{ \partial r_{\alpha 1}}  \Bigr) \, .\nonumber\\
 \label{pairprob}
\eea  
The solution is
\be
P^{(2)}_2(\rr_1,\rr_2) =
\frac{1}{Z_2}
 \exp\Bigl(-\frac{\psi(\rr_1,\rr_2)+\frac{\tau}{2 \gamma}\sum_{\alpha=1}^d
[ \frac{\partial }{\partial r_\alpha} \psi(\rr_1,\rr_2) )]^2  - D\gamma \ln\det \Gamma(\rr_1,\rr_2) }{T_s}   \Bigr)
 \label{pairproc}
\ee
where $\psi(\rr_1,\rr_2)=u(\rr_1)+u(\rr_2)+w(\rr_1-\rr_2)$ and
$\det \Gamma$ is the determinant associated with the $2d\times 2d$ matrix whose elements are
\be
\Gamma_{\alpha \beta }(\rr_i,\rr_j)= \delta_{\alpha\beta}   \delta_{ij}+\frac{\tau}{\gamma}
\frac{\partial^2  \psi(\rr_i,\rr_j) }{\partial r_{\alpha i}\partial r_{\beta j}} 
\ee
with $i,j=1,2$.
The form of eq. \eqref{pairproc} suggests the idea of introducing an effective potential to describe the interaction experienced by the particles
when subjected to colored noise. The effective potential can simplify the description, make the 
analysis more transparent and avoid the difficulty of evaluating the inverse matrix $\Gamma$
when the system comprises a large number of particles.
However, we must explore the validity of such a method since it involves an approximate treatment of the interactions
when $N\geq3$.
 Let us begin with the simplest case
of just two particles  free to move on a line  in the absence of external fields and write the pair distribution. To this purpose let us consider
the $2\times2$ matrix $\Gamma_{ij}$:
$$
\Gamma^{(2)}=
\left(\begin{array}{ccccccc}
1+\frac{\tau}{\gamma} w_{11}(x_1-x_2)&  -\frac{\tau}{\gamma}w_{11}(x_1-x_2)
\\ -\frac{\tau}{\gamma}w_{11}(x_1-x_2)   & 1+\frac{\tau}{\gamma} w_{11}(x_1-x_2)
\end{array}\right)
$$
with $w_{11}=\frac{d^2w(x_1-x_2)}{d x_1^2}$ .
The resulting two particles distribution function $P_2^{(2)}$  has the form:
\be
P_2^{(2)}(x_1-x_2)=\frac{1}{Z_2}\exp\Bigl( -\frac{\phi(x_2-x_1)}{T_s}\Bigl)
\label{effectiveone}
\ee
Thus,
we can define, apart from a constant, the following pair
effective potential  by taking the logarithm of $P_2$
 \be
  \phi(x_1-x_2)=w(x_1-x_2)+   (\frac{\tau}{\gamma})[  w_1(x_1-x_2)  ]^2- D\gamma \ln (1+2 (\frac{\tau}{\gamma})  w_{11}(x_1-x_2))
   \ee
with $w_1=\frac{dw(x_1-x_2)}{d x_1}$.

The above result can be generalized in the case of higher dimensionality. 
The pair correlation function  in $d$ dimensions for a two-particle system interacting via a central potential $w(r)$, in the absence of external potentials,  reads:
   \be
  g(r)=   \exp \left(- \frac{ w(r)   +\frac{\tau}{ \gamma} [ w'(r)]^2 
   - T_s\ln [(1+ 2 \frac{\tau}{ \gamma}  w''(r)  )(1+2\frac{\tau}{ \gamma} \frac{w'(r)}{r})^{d-1}]  } {T_s}   \right) \ee
   where the primes mean derivative with respect to the separation $r$.
   Thus the  effective pair potential reads
$  \phi(r)=-T_s\ln g(r)$.  
  Let us remark that the effective potential, being derived in the low-density limit, does not account for the three body terms
  which instead are present  if one considers formula \eqref{probabilitydistr} with $N>2$.

In all these cases the dependence of $\phi$  on the effective temperature is quite interesting because as $D$  increases the effective potential 
displays a deeper and deeper potential well.  
Hereafter, we shall adopt the following unit system: lengths are expressed in terms of the molecular length $\sigma$,
time in terms of the unit time $\gamma^{-1}$ and the unit mass is set equal to 1.

\begin{figure}[h]
\centering
  \includegraphics[height=10cm]{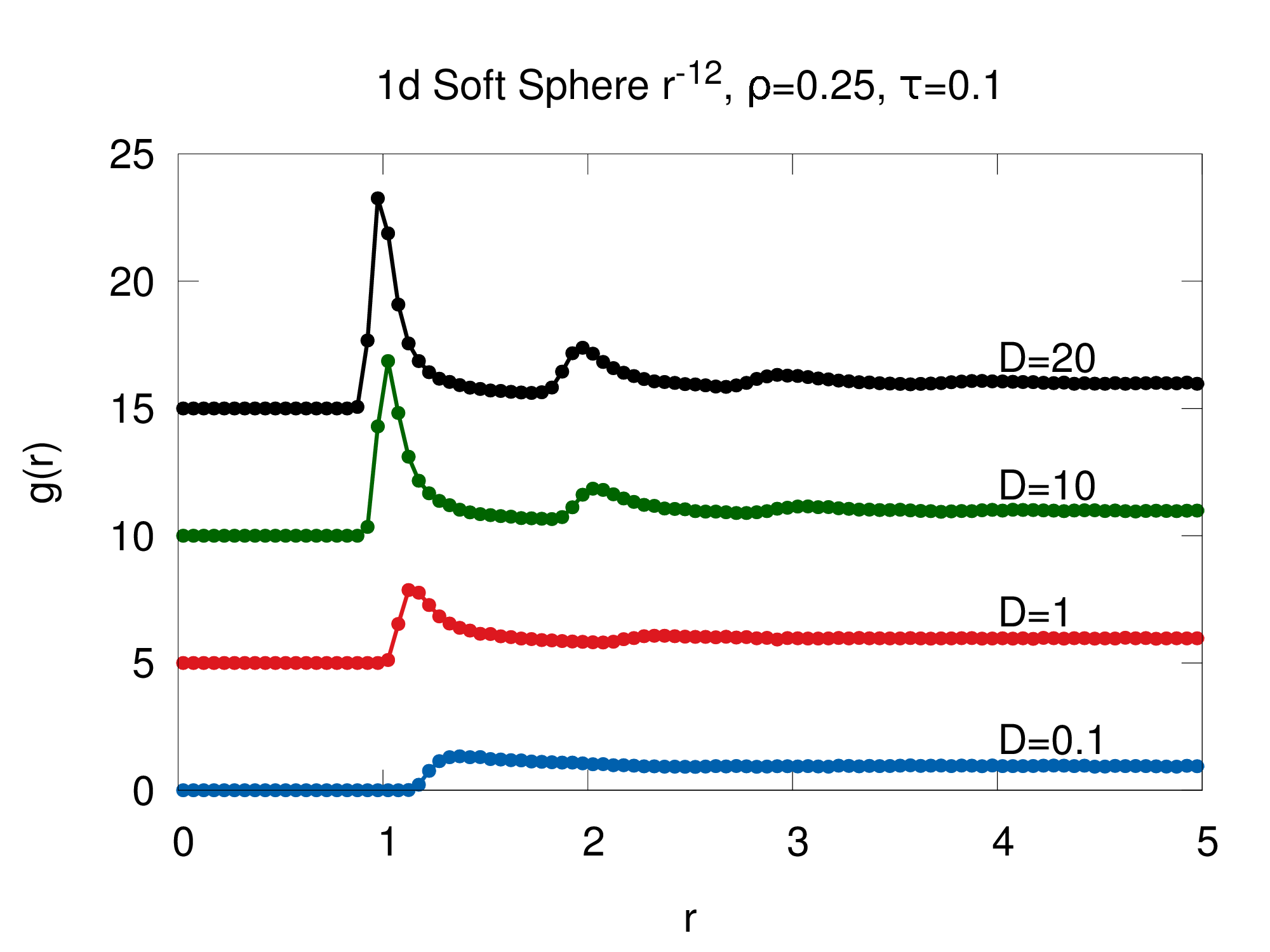}
  \caption{Results for the pair correlation function  obtained via numerical simulation of a 1 dimensional system. 
  Density $\rho$ is expressed in units $\sigma^{-1}$, $\tau$ in units $\gamma^{-1}$ and $D$ in units $T_s/\gamma$. }
  \label{peakversusD}
\end{figure}

In particular, we find that for a pair potential $w(x)=w_0(\frac{\sigma}{x})^{12}$, in order to observe an attractive region it is 
necessary to have $2 D\tau/\sigma^2>1$.
This is shown in Fig.   \ref{peakversusD}   where we display the pair correlation function obtained by numerical simulations  of a system of repulsive particles
in one dimension for different values of $D$, average density  $\rho\sigma=0.25$  
and for  $\tau=0.1$.
The various curves correspond to different values of $D$  and one can see that the height of the peak increases with $D$ because the  effective attraction increases.
Such an effective  attraction in a system  where only repulsive interparticle forces are in action  is
due to a dynamical  mechanism. 
It can be understood  as follows:   the friction that a particle
experiences with the background fluid is enhanced by the presence of surrounding particles so that their mobility
decreases.
Being less mobile the particle tends to spend more time in configurations where it is closer to other particles and 
one interprets this situation as an effective attraction \cite{cates2013active,cates2014motility,barre2014motility}.



\begin{figure}[h]
\centering
  \includegraphics[height=12cm]{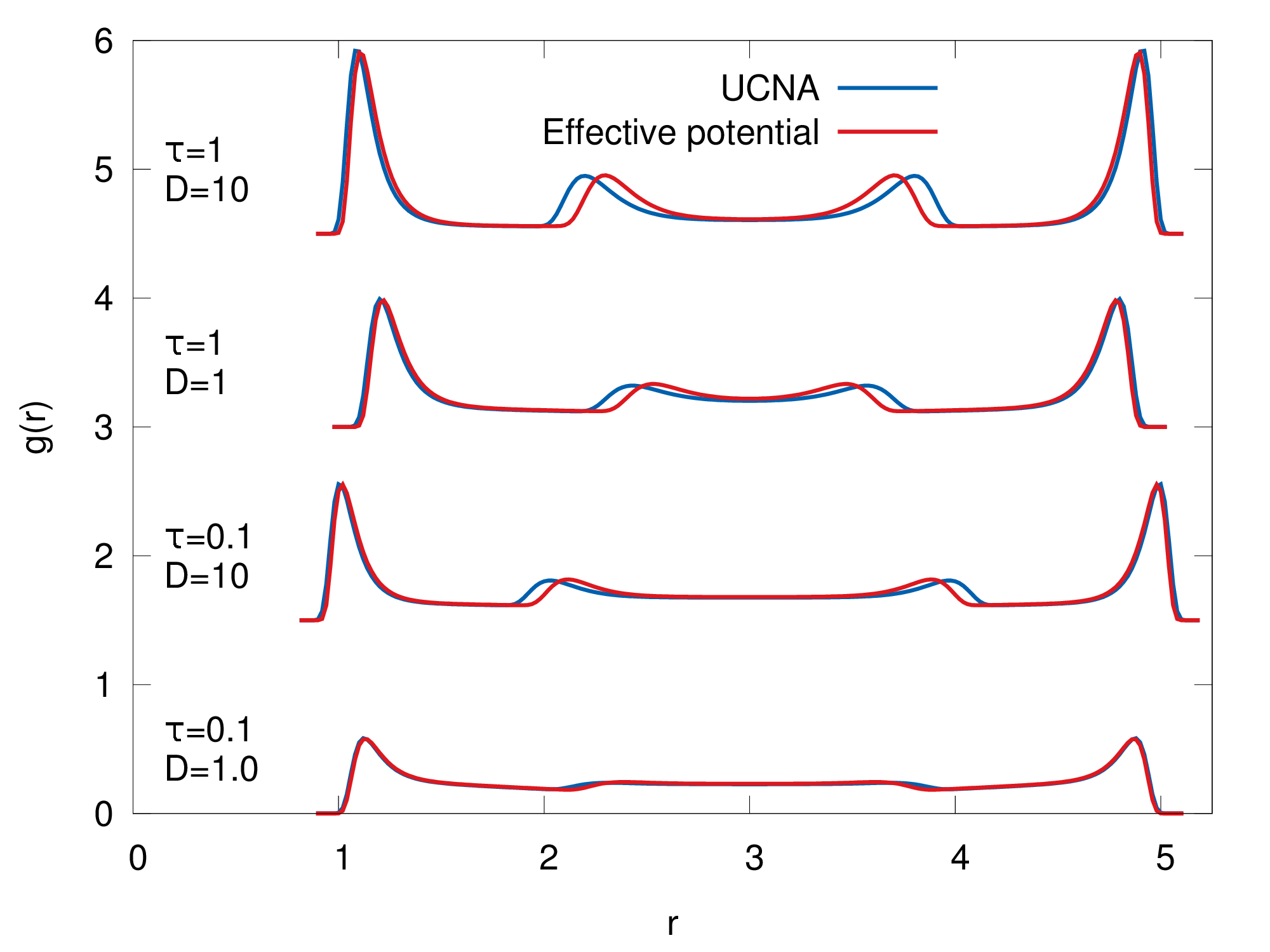}
  \caption{Normalized single particle probability distribution function for a one dimensional system two interacting particles confined between two repulsive walls
  at $x=0$ and $x=6$. The wall repulsive potential is $u(x)=u_0(\frac{\sigma}{x})^{12}$ at each wall and $w(x_1-x_2)=w_0(\frac{\sigma}{(x_1-x_2)})^{12}$  between the particles with $u_0=w_0=1$. The curves refer to  $\tau=0.1,1.$ and $D=1,10$ expressed
  in the same units as in fig. \ref{peakversusD}.
   The agreement between the Gaussian colored noise result and the effective potential is  poor as far as the second peak is concerned:
   its position is shifted towards larger distances from the wall, when the  persistence time increases.  }
  \label{comparisoneffective}
\end{figure}

In  spite of the fact that the MUCNA  N-particle distribution function is known it is difficult to apply it to large systems
because of the many-body nature of the interactions. On the other hand,
if one could replace   the complicated MUCNA potential  by an effective 
pairwise potential it would possible to use all the machinery employed in the study 
of molecular fluids. Under this hypothesis  one could, for instance, use methods such as the density functional method,
define a Helmholtz intrinsic free energy, or utilize the integral equations method and  greatly simplify the study the phase behavior of the model.  To analyze its validity, 
we use a simple example: consider an interacting two-particle system subjected to the action of an external potential $u(x)$ whose distribution function is given by
\bea
 && P_2^{(2)}(x_1,x_2)=
   \frac{1}{Z_2}   \, \exp \Bigl(- \frac{w(x_1,x_2)+(u(x_1)+u(x_2)  }{T_s} \Bigr) \nonumber\\
 &&
 \times\exp\Bigl(  -\frac{\tau}{2 \gamma T_s} [\frac{\partial}{\partial x_1} (w(x_1,x_2)+u(x_1))]^2 -\frac{\tau}{2 \gamma T_s} [\frac{\partial}{\partial x_2} (w(x_1,x_2)+u(x_2))]^2
 \nonumber\\
 &&
 + \ln \det \Gamma(x_1,x_2)\Bigr)
    \eea
with
    \bea
  \det \Gamma(x_1,x_2)&=&1+\frac{\tau}{\gamma} \Bigl(
  2 \frac{\partial^2 w(x_1,x_2)}{\partial x_1^2}+  \frac{\partial^2 u(x_1)}{\partial x_1^2} + \frac{\partial^2 u(x_2)}{\partial x_2^2}  
  \Bigr) \nonumber\\
  &&
  +(\frac{\tau}{\gamma})^2 \Big(\frac{\partial^2 w(x_1,x_2)}{\partial x_1^2} [  \frac{\partial^2 u(x_1)}{\partial x_1^2}
  + \frac{\partial^2 u(x_2)}{\partial x_2^2}   ]+ \frac{\partial^2 u(x_1)}{\partial x_1^2}
  \frac{\partial^2 u(x_2)}{\partial x_2^2}  \Bigr) .\nonumber\\   \eea
 In the spirit of the effective potential idea we introduce the following superposition approximation
  \bea
  && P_2^{approx}(x_1,x_2)\approx  \frac{1}{\ Z_2^{approx}}   \, \exp \Bigl(- \frac{w(x_1,x_2)+u(x_1)+u(x_2)  }{T_s} \Bigr) \nonumber\\
 &&
 \times\exp\Bigl(  -\frac{\tau}{2 \gamma T_s} [\frac{\partial}{\partial x_1} w(x_1,x_2)]^2 -\frac{\tau}{2 \gamma T_s} [\frac{\partial}{\partial x_2} w(x_1,x_2)]^2-
  \frac{\tau}{2 \gamma T_s} [\frac{\partial}{\partial x_1} u(x_1)]^2 -\frac{\tau}{2 \gamma T_s} [\frac{\partial}{\partial x_2} u(x_2)]^2 \nonumber\\
  &&
+ \ln \det \Gamma^{approx}(x_1,x_2)\Bigr) 
\label{p2approx}
\eea
  with
 \be
 \ln \det \Gamma^{approx}(x_1,x_2)\approx 
 \ln \Bigl(1+2 (\frac{\tau}{\gamma})\frac{\partial^2 w(x_1,x_2)}{\partial x_1^2} \Bigr)+
\ln \Bigl( 1+(\frac{\tau}{\gamma})   \frac{\partial^2 u(x_1)}{\partial x_1^2}\Bigr) + \ln \Bigl( 1+(\frac{\tau}{\gamma})   \frac{\partial^2 u(x_2)}{\partial x_2^2}  \Bigr) .  \ee
Clearly the exact formula and the approximation for the determinant differ beyond the linear order in $\tau/\gamma$, but perhaps the largest discrepancy occurs
in the presence of cross terms of linear order in $\tau/\gamma$, such as $w(x_1,x_2)u(x_1)$ in the interaction potential .
A test can be performed 
by comparing the probability density profile $P^{(1)}(x_1)$ obtained by integrating over the coordinate $x_2$ the exact distribution and the
one obtained by  applying the same procedure to the approximate distribution \eqref{p2approx}.
The comparison, displayed in Fig. \ref{comparisoneffective}, reveals the presence of a systematic shift of the second peak of the approximate distribution
towards larger distances from the  confining wall, as if the total effective force were more repulsive.



\begin{figure}[h]
\centering
  \includegraphics[height=10cm]{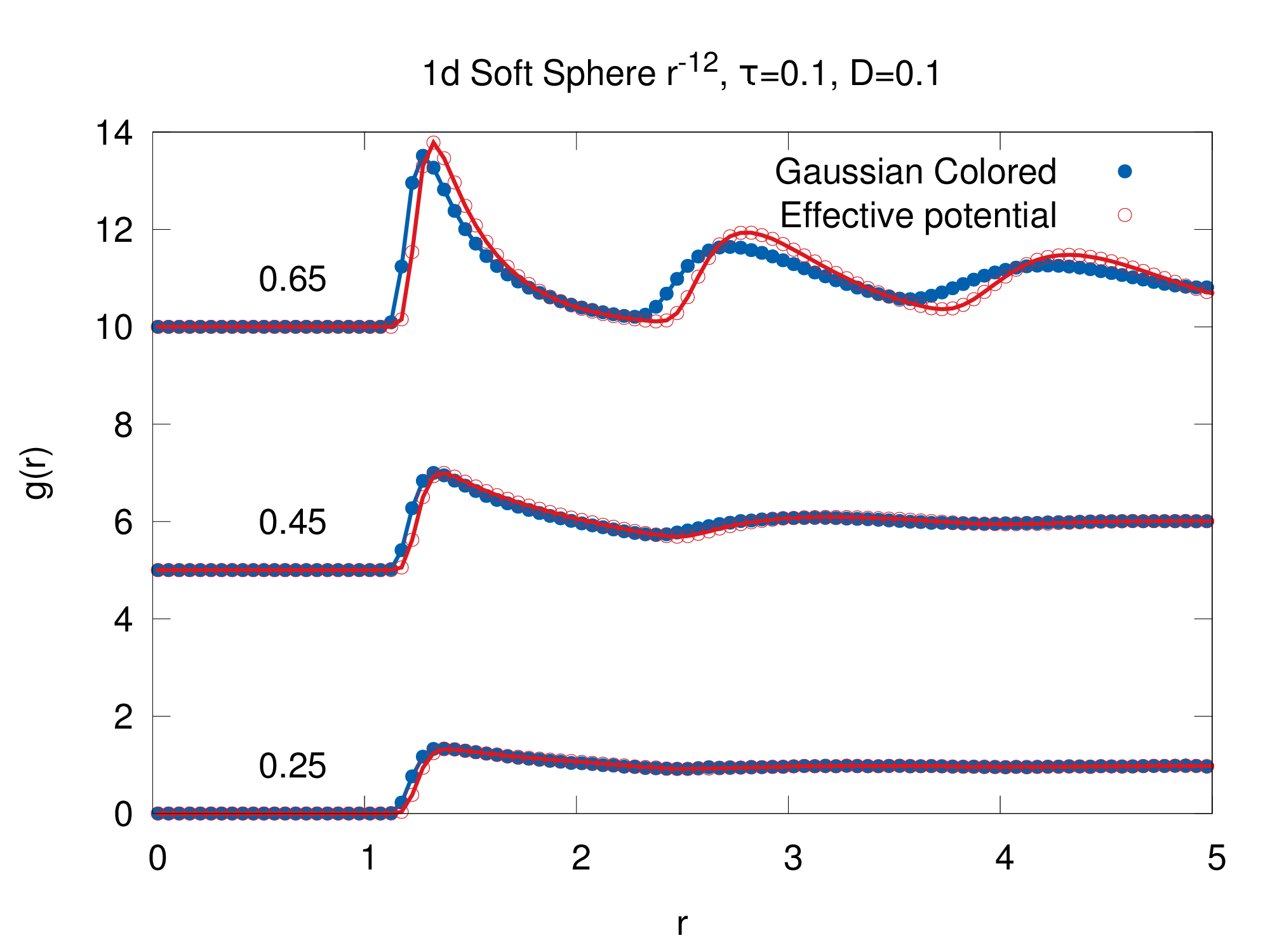}
  \caption{Numerical simulation of a 1 dimensional system.
  Pair correlation for  $\rho\sigma=0.25,0.45,0.65$  for fixed $\tau=0.1$ and  $D=0.1$ , expressed
  in the same units as in fig. \ref{peakversusD}.
 The height of the peak increases with density. The agreement between the Gaussian colored noise result and the effective potential is moderately good.  }
  \label{fgr:onedimension1}
\end{figure}

\begin{figure}[h]
\centering
  \includegraphics[height=10cm]{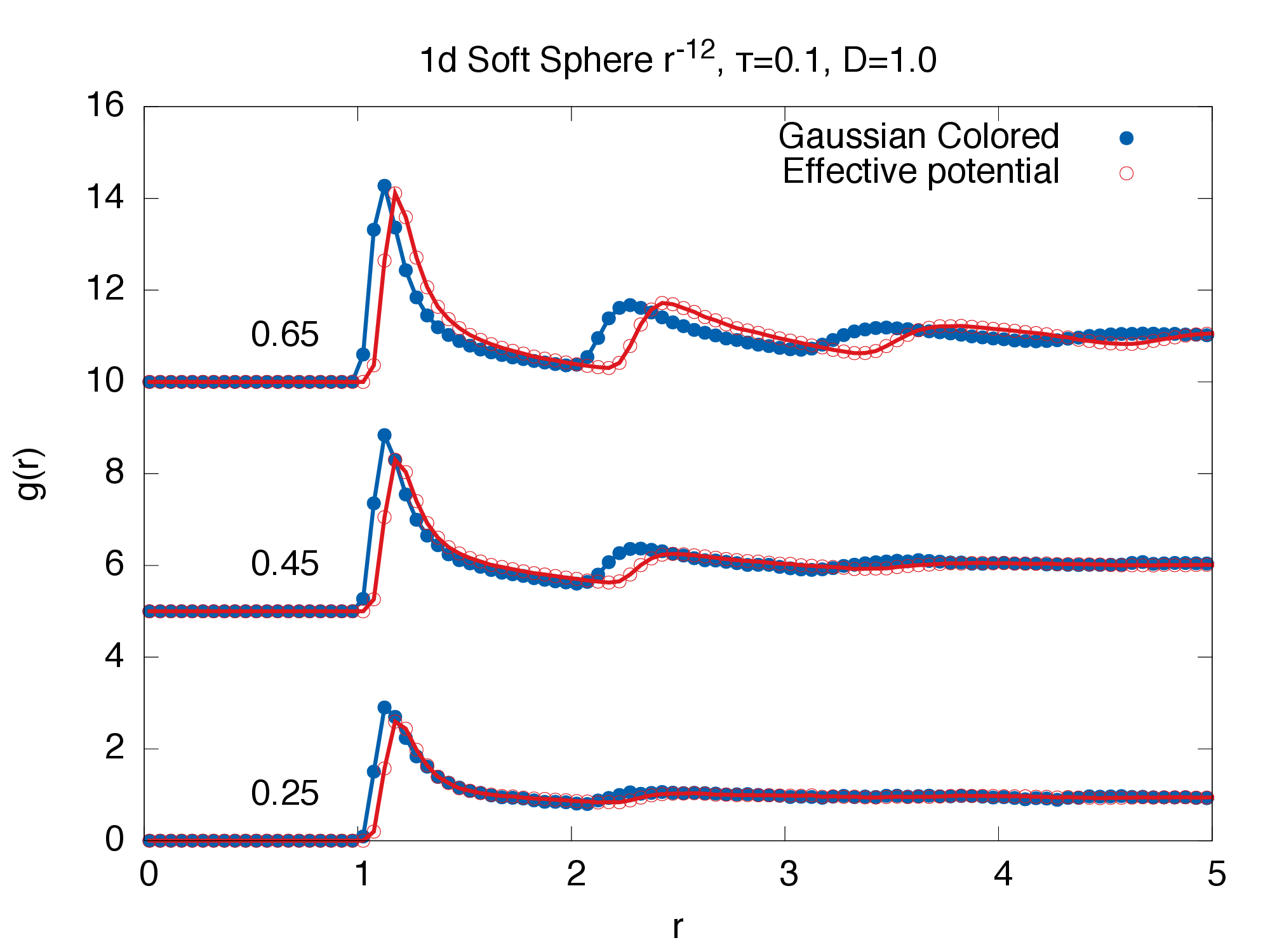}
  \caption{Numerical simulation of a 1 dimensional system. Pair correlation for  $\rho\sigma=0.25,0.45,0.65$  for fixed $\tau=0.1$ and  $D=1$, expressed
  in the same units as in fig. \ref{peakversusD}.
   The agreement between the Gaussian colored noise result and the effective potential is rather poor due to the relatively large value of the persistence time.  }
  \label{fgr:onedimension2}
\end{figure}

\begin{figure}[h]
\centering
  \includegraphics[height=10cm]{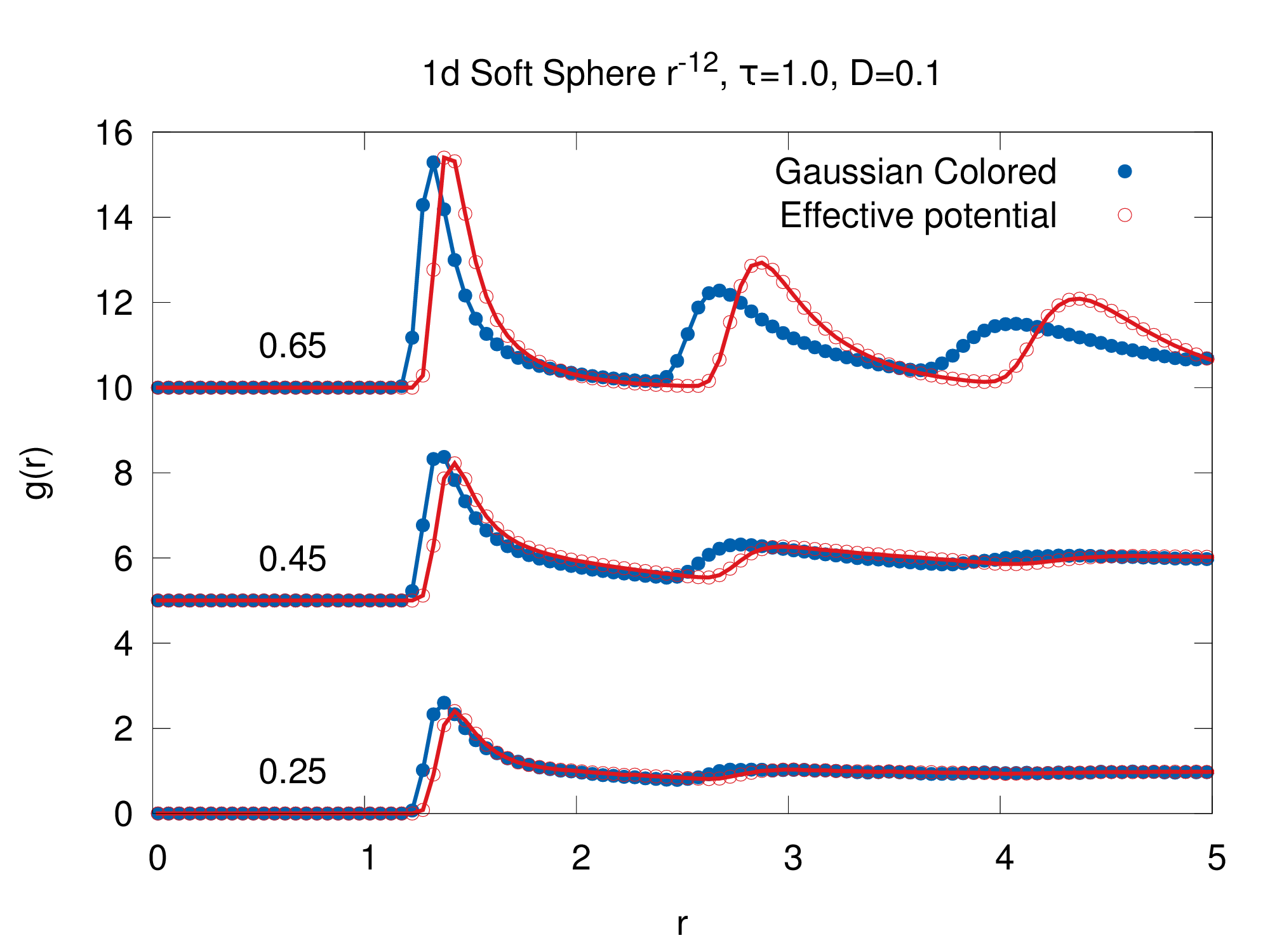}
  \caption{Numerical simulation of a 1 dimensional system. Pair correlation for  $\rho\sigma=0.25,0.45,0.65$  for fixed $\tau=1.$ and  $D=0.1$
   The agreement between the Gaussian colored noise result and the effective potential is rather poor due to the relatively large value of the persistence time.  }
  \label{fgr:onedimension3}
\end{figure}

\begin{figure}[h]
\centering
  \includegraphics[height=10cm]{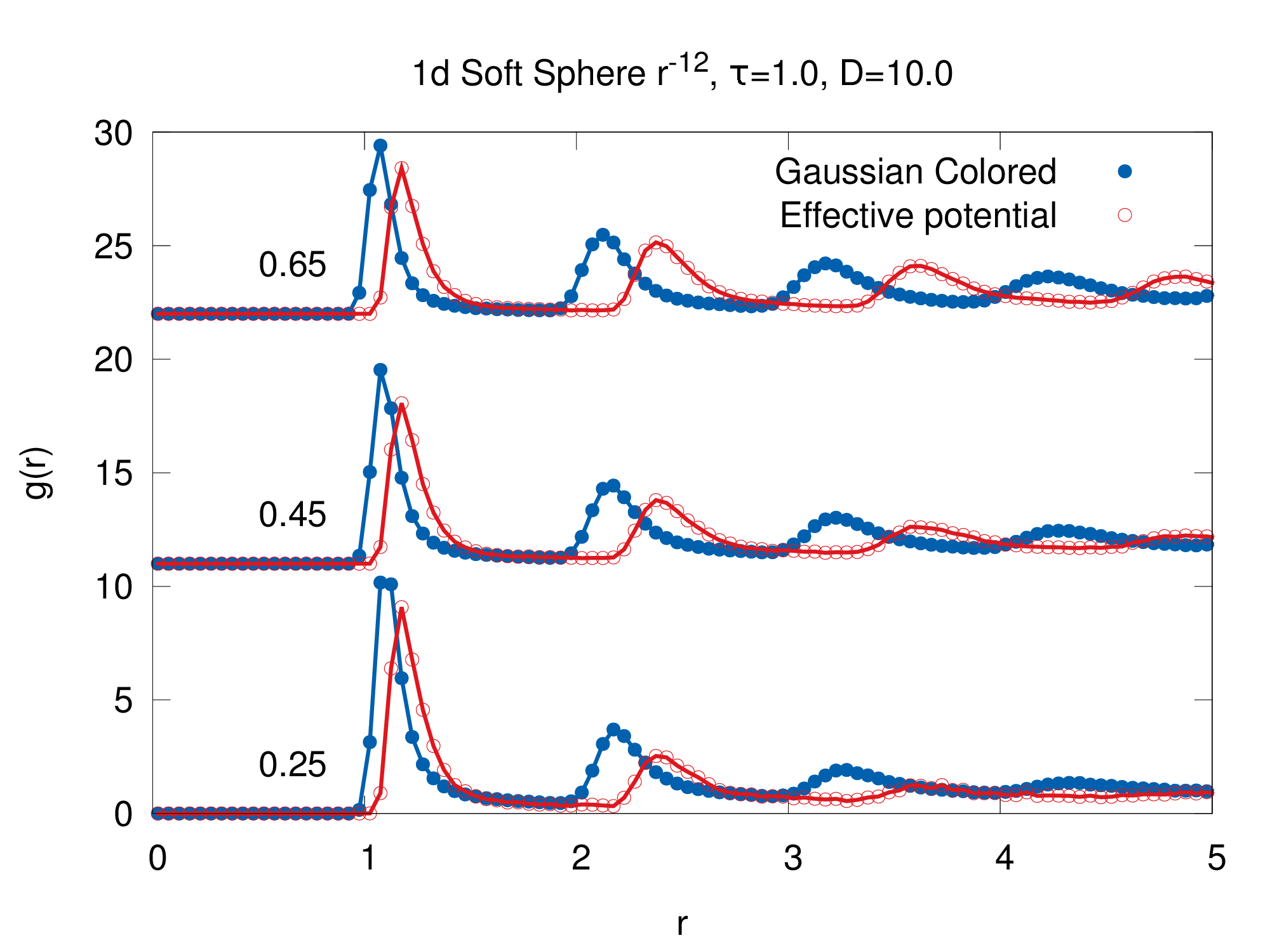}
  \caption{Numerical simulation of a 1 dimensional system. Pair correlation for  $\rho\sigma=0.25,0.45,0.65$  for fixed $\tau=1.$ and  $D=10.$
   The agreement between the Gaussian colored noise result and the effective potential is rather poor due to the relatively large value of the persistence time.  }
  \label{fgr:onedimension5}
\end{figure}

  We turn, now, to consider a many-particle systems and perform a similar comparison. 
We have simulated the system described by eqs.  \eqref{effective_langevin} - \eqref{exp_correlation} with $N=1000$ and computed numerically the pair correlation function.
In order to check the effective potential approximation we also performed simulations of the over-damped Langevin equation
with white noise for particles in one dimension subjected to interactions given by $\phi(x_i-x_j)$  .
The corresponding  results are shown in figs. \ref{fgr:onedimension1}-\ref{fgr:onedimension5}.
One observes in figs. \ref{fgr:onedimension1}    and \ref{fgr:onedimension2}  that at moderately low values of  the persistence time, $\tau=0.1$, the discrepancy between the effective potential approximation and the
full colored noise result is not too large even at large densities, although there is a systematic shift of the peaks in the effective potential towards
larger values of the distance.
The situation at values of $\tau$ ten times larger, $\tau=1$, and $D=0.1$ and $10$ is remarkably worse and the peaks of the effective theory 
display a much larger  
shift as illustrated by figs. \ref{fgr:onedimension3}-\ref{fgr:onedimension5}. 
Such a shift is determined by the approximate treatment of the three-body repulsive term which appears in eq.  \eqref{hamilt1}
which becomes more relevant as the density and $\tau$ increase.
These findings pose some limits to the possibility of obtaining reliable
results by employing the effective potential approximation for values of the  persistence time too large.

\subsection{Van der Waals free energy}

 We use, now, a van der Waals (vdW) argument to estimate the free energy for  the present model in $d$-dimension when the bare potential is
 of the form $w(r)=w_0(\frac{\sigma}{r})^\alpha$
  and identify the following repulsive  contribution in the effective potential:
  \be
\phi_{rep}(\tilde r)= 
 w_0(\frac{\sigma}{r})^\alpha+ \frac{\tau }{\gamma}\frac{\alpha^2}{ \sigma^2}w_0^2 (\frac{\sigma}{r})^{2\alpha+2}
\ee

whereas the  attractive contribution is:
\be
\phi_{attr}(r)=  - T_s\ln\Bigl(\Bigl[1+2 \alpha(\alpha+1)    \frac{ w_0\tau}{\gamma \sigma^2}  (\frac{\sigma}
{ r})^{\alpha+2} \Bigr] \Bigl [1- 2  \alpha \frac{w_0 \tau}{\gamma \sigma^2}(\frac{\sigma} { r})^{\alpha+2} \Bigr]^{d-1}\Bigl) .
 \label{attractive}
   \ee  
Thus for $w_0 \frac{\tau}{\gamma \sigma^2}<<1$ the system reduces to a system of passive soft repulsive spheres. 
Using a standard procedure
we represent the free energy as the sum of a  repulsive contribution evaluated in the 
local density approximation plus a non local mean-field attractive term:
\bea
&&
\Fu [\rho^{(1)} ] = 
T_s \int d^dr \rho  (\rr) \Bigl[\ln \Bigl(
\frac{\rho^{(1)} (\rr)}  {   1-  b\rho^{(1) }(\rr) }
\Bigr)-1\Bigr]   
  +\frac{1}{2} \iint_{|\rr-\rr'|>R_b}
  d^d r  d^d r' \,  \phi_{attr}(\rr-\rr')\rho^{(1)}(\rr) \rho(\rr') \nonumber \\
\label{funct}
\eea
where $b=\omega_d R^d$, 
 $\omega_d=1, \pi/2,2\pi/3$ for $d=1,2,3$, respectively and
the effective hard-sphere diameter, $R_b$ is given by the Barker-Henderson formula:
\be
R_b=\int_0^\infty dr (1-e^{-\phi_{rep}(\tilde r)/T_s}) .
\ee
Following the standard vdW approach we may represent the pressure associated with the functional \eqref{funct} as:
   \be
p=
 \frac{ T_s \,\rho}{1-b \rho}-a\rho^2 ,
\label{vdW}
\ee   
where the value of the coefficient $a$ is determined by the strength of the effective attractive interaction:
\be 
a =- \int_{R_b}^\infty d^d  r \, \phi_{attr}(r) .
 \ee  
From the form of the effective attractive potential \eqref{attractive} one sees that $a$ is an increasing function of the non dimensional parameter $\frac{w_0 \tau}{\gamma \sigma^2}$ and of $T_s$.
The latter feature determines a remarkable difference with respect to passive fluids: in that case the vdW pressure 
is the sum of an entropic term proportional to the temperature and of a negative temperature independent enthalpic term.
In active fluids, on the contrary,  the coefficient $a$ increases roughly linearly  with $T_s$ as the entropic term does.
As a result, $T_s$, since $a/T_s$ is nearly constant, does not play a major role in determining the critical parameters of the active model.
The effective  attraction among the particles
has its origin in the reduction of their mobility, reflected by the presence of an effective friction  $\Gamma_{ij}$ in \eqref{stochasticxi}
 which increases with increasing density.
 As a result the particles tend to accumulate where their density is higher and move more slowly and possibly lead to the mobility induced phase separation phenomenon.

In analogy with the vdW model of passive fluids one can determine the critical parameters, where a second order transition would
take place,
by solving simultaneously the  equations 
$\frac{\partial p}{\partial V}=0$ and $\frac{\partial^2 P}{\partial V^2}=0$
with the result:
\bea
\frac{a(\tau_c)}{T_s}=\frac{27}{8 } b  \,, \,\,\,\,\,\,  
\rho_c=\frac{1}{3 b}
\label{valuescrit}
\eea
Thus,  in order to have phase separation,  the persistence time, $\tau$, must exceed the critical value
$\tau_c$, implicitly given  by the first of equations \eqref{valuescrit}.
However, the numerical investigation of the phase separation of
Gaussian-colored noise driven particles is still in progress and so far it has not revealed a  clear phase separation in the absence of attraction.



\section{Conclusions}

We  discussed the properties of a newly introduced model describing an active fluid, consisting of an 
 an assembly  of repulsive soft spheres subject to over-damped dynamics and driven by a colored noise.
 Although  within the MUCNA  is possible to get an analytical expression for the many-particle distribution function it is difficult to make progress without resorting to approximations. The 
  effective potential hypothesis represents a practical possibility since it reduces the many-body potential  to a pairwise additive potential where one can use the
  standard tools of statistical mechanics.
 To test the hypothesis, we performed two checks.
   In the first we employed a toy model consisting of just two particles in an external field and performed explicitly the analytic calculations.
In the second check, we compared by numerical Brownian simulation  the properties of a one-dimensional system of soft repulsive spheres
subjected to colored noise against the corresponding properties of a system of particles interacting via 
the effective potential.
It is found that for values of the persistence time, not too large the effective potential approximation is reliable.

\section{Acknowledgements}
C. M. acknowledges support from the European Research
Council under the European Union?s Seventh Framework programme (FP7/2007-2013)/ERC Grant agreement No. 307940.
M.P. acknowledges support from NSF-DMR-1305184.


\appendix

\section{Derivation of the UCNA equation  by multiple time-scale analysis}

\label{multipletimescale}

In this appendix we derive the UCNA approximation by a multiple time-scale analysis
following the same method employed in ref. \cite{marconi2006nonequilibrium,marini2007}. It allows to derive in a systematic fashion
the configurational  Smoluchowski equation  from the Kramers equation  via the elimination of the velocity degrees of freedom.
To achieve this goal one introduces fast and slow time-scale  variables for the independent variable, and subsequently treats these variables, fast and slow, as if they are independent.  The solution is first expressed as a function of these different time scales and subsequently  these new independent variables are used to remove secular terms in the resulting perturbation theory.
 Physically speaking, the fast scale  corresponds
to the time interval necessary to the velocities of the
particles to relax to configurations consistent with the 
 values imposed by the vanishing of the currents. The slow time scale is much longer and corresponds
to the time necessary to the positions of the particles to relax towards the stationary configuration.

It is convenient to work with non dimensional quantities and  introduce 
the following variables: 
\begin{equation}
\bar t\equiv t\frac{v_T}{l}, \qquad V\equiv\frac{v}{v_T}, 
\qquad X\equiv\frac{x}{l},\qquad \Gamma=\gamma\frac{l}{v_T} ,\qquad \zeta=\frac{l}{\tau v_T}
\label{adim1}
\end{equation}

\begin{equation}
K(X)\equiv\frac{l F(x)}{m v_T^2} ,\qquad v_T=\sqrt{D/\tau}\label{adim2}
\end{equation}
where  $l$ is a typical length of the problem, such as the size of the particles and $\zeta$ plays the role of a non dimensional friction.

It is clear that
if $\zeta>>1$ particles lose memory of their initial
velocities after a time span which is of the order 
of the time constant $\tau$ so that the velocity distribution 
soon becomes stationary.
We shall assume that $\gamma \tau$ stays finite when $\tau\to 0$.
In this limit the Smoluchowski description of a system of non interacting
particles, which takes into account
only the configurational degrees of freedom, turns out to be adequate.
However, for intermediate values of $\tau$ the velocity may play a role.
 The question is how do we recover the UCNA  starting from a description
in the larger space $x,v$ ?
We rewrite  Kramers' evolution equation for 
the phase-space distribution function
using  (\ref{adim1}-\ref{adim2}) 
as:
\begin{equation}
\frac{\partial \tilde f(X,V,\bar t)}{\partial \bar t} +V \frac{\partial }{\partial X} \tilde f(X,V,\bar t) 
+ \frac{\zeta}{\Gamma}K(X,\bar t) \frac{\partial }{\partial V} \tilde f(X,V,\bar t)
=\zeta L_{FP}\tilde f(X,V,\bar t)
\label{kramers0}
\end{equation}
having introduced the ``Fokker-Planck'' operator
\be
L_{FP}\tilde f(X,V,\bar t) =\frac{\partial}{\partial V}\Bigl[
\frac{\partial }{\partial V }+s(X) V\Bigl]  \tilde f(X,V,\bar t)
\label{fokkerp}
\ee
with 
$$
s(X)=1-\frac{1}{\Gamma \zeta}\frac{d K}{dX}
$$
whose eigenfunctions are
$$H_{\nu}(X,V)
= (-1)^{\nu}   \frac{1}{\sqrt{2\pi}} s^{-(\nu-1)/2}
 \frac{\partial^{\nu}}{\partial V^{\nu}} \exp(-\frac{s(X)}{2}V^2)
$$
and have non positive  eigenvalues $\nu=0,-s,-2 s, .., -\nu s$.
Notice that as stated above we treat $\zeta/\Gamma$ as a quantity of order 1.
Solutions of eq. (\ref{kramers0}),
where the velocity dependence of the distribution function 
is separated, can be written as:
\begin{equation}
\tilde f(X,V,\bar t) \equiv \sum_{\nu=0}^{\infty}\phi_{\nu}(X,\bar t) H_{\nu}(X,V).
\label{expansion}
\end{equation}

In the multiple time-scale analysis one determines the temporal evolution
of the distribution function $\tilde f(X,V,\bar t)$ in the regime
$\zeta>>1$, by means of a perturbative method. 
In order to construct the solution one replaces the single
physical time scale, $\bar t$, by a series of auxiliary time scales
($\bar t_0,\bar t_1,..,\bar t_n$) which are related to the original variable
by the relations $\bar t_n=\zeta^{-n}\bar t$. Also the original
time-dependent function, $\tilde f(X,V,\bar t)$,  
is replaced by an auxiliary function,$\tilde f_a(X,V,\bar t_0,\bar t_1,..)$,  
that depends on the $\bar t_n$,  treated as independent variables.
Once the equations corresponding to the various orders have been 
determined, one returns to the original time variable and to the
original distribution.

One begins by replacing the time derivative
with respect to $\bar t$ by a sum of partial derivatives:
\begin{equation}
\frac{\partial}{\partial \bar t}=\frac{\partial}{\partial \bar t_0}
+\frac{1}{\zeta} \frac{\partial}{\partial \bar t_1}
+\frac{1}{\zeta^2} \frac{\partial}{\partial \bar t_2}+..
\label{mult}
\end{equation}
First, the  function,$\tilde P(X,V,\bar t_0,\bar t_1,..)$
is expanded as a series of $\zeta^{-1}$
\begin{equation} 
\tilde P(X,V,\bar t_0,\bar t_1,\bar t_2,..)=
\sum_{s=0}^{\infty} \frac{1}{\zeta^s} 
\sum_{\nu=0}^{\infty}  
\psi_{s \nu}(X,\bar t_0,\bar t_1,\bar t_2,..)H_{\nu}(X,V)
\label{pn}
\end{equation}

One substitutes, now, the time derivative~(\ref{mult})
and expression~\eqref{pn} into eq. 
(\ref{kramers0}) and identifying terms 
of the same order in $\zeta^{-1}$ in the equations
one obtains a hierarchy of relations
between the amplitudes $\psi_{s \nu}$. 
To order $\zeta^0$ one finds:
\begin{equation}
L_{FP} \Bigr[\sum_\nu \psi_{0\nu}H_\nu\Bigr]=0
\label{g0}
\end{equation} 
and concludes that only the amplitude $\psi_{00}$ is non-zero. 

Next, we consider terms of order  $\zeta^{-1}$ and write:
\begin{equation}
L_{FP} \Bigr[ \psi_{11}H_1+\psi_{12}H_2 + \psi_{13}H_3 \Bigr]=
\frac{\partial \psi_{00}}{\partial \bar t_0}H_0+
\Bigl(V \frac{\partial }{\partial X} 
+ \frac{\zeta}{\Gamma}K(X,\bar t) \frac{\partial }{\partial V}\Bigl)
 H_0(X,V)\psi_{00}\label{g1}
\end{equation} 
After some straightforward calculations and
 equating the coefficients multiplying the same $H_{\nu}$ 
we find:
\begin{equation}
\frac{\partial \psi_{00}}{\partial \bar t_0}=0
\label{psi0ta}
\end{equation} 
and

\begin{equation}
\psi_{11}=-s^{-3/2}\Bigl( \frac{\partial }{\partial X }-\frac{s'}{s}-
s\frac{\zeta}{\Gamma}K \Bigl)\psi_{00}
\label{psi0t}
\end{equation} 
where $s'=d s/dx$.
According to \eqref{psi0ta}
the amplitude $\psi_{00}$ is constant with respect to  $\bar t_0$
and so is $\psi_{11}$
being a functional of $\psi_{00}$.
The remaining amplitudes $\psi_{1k}=0$ are zero for all $k>1$
with the exception of
\begin{equation}
\psi_{13}=\frac{1}{6} s^{-5/2} s'\psi_{00}\
\label{psi0txx}
\end{equation}

The equations of order $\zeta^{-2}$ give 
the  conditions:
\begin{equation}
\frac{\partial \psi_{00}}{\partial \bar t_1}=
-  \frac{\partial }{\partial X } (\frac{\psi_{11}}{s^{1/2}})
\label{psi0t1}
\end{equation}

If we truncate the expansion to second order and
collect together the various terms 
and employing eq.~(\ref{mult})  to  restore the original time
variable $\bar t$  we  obtain the following evolution equation:
\begin{equation} 
\frac{\partial \psi_{00}}{\partial \bar t}=
\frac{1}{\zeta} \frac{\partial }{\partial X }\Bigl[\frac{1}{s}\Bigl( \frac{\partial }{\partial X }(\frac{\psi_{00}}{s})-
\frac{\zeta}{\Gamma}K\psi_{00} \Bigl) \Bigl]
\label{zero0}
\end{equation}
Now we can return to the original dimensional variables:

\begin{equation} 
\frac{\partial P(x,t)}{\partial  t}=
D \frac{\partial }{\partial x }\Bigl\{\frac{1}{1-\frac{\tau}{\gamma} F'(x)}\Bigl[ \frac{\partial }{\partial x }\Bigl(\frac{P(x,t)}{1-
\frac{\tau}{\gamma} F'(x)} \Bigl)-
\frac{1}{D\gamma}F(x) P(x,t)  \Bigl]\Bigr\}
\label{zero0b}
\end{equation}
Equivalently, such a  result  would have followed by starting form the effective Langevin equation: 
\bea
\frac{d x}{d t} =  \frac{1}{\gamma }  \frac{F(x)}{1-\frac{\tau}{\gamma}  F'(x)} 
+\frac{D^{1/2} }{1-\frac{\tau}{\gamma} '(x)}\xi^{w}(t)
\eea
which displays a space dependent friction and a space dependent noise.

Clearly the stationary configurational distribution associated with \eqref{zero0b} is:
\be
P(x)= {\cal N} (1+ \frac{\tau}{ \gamma} \frac{d^2 u}{d x^2} )\exp\Bigl[
-\frac{1}{D\gamma}\Bigl (u(x)+ \frac{\tau}{2 \gamma}
(\frac{d u}{d x})^2 
\Bigl)\Bigl]
\ee
where we introduced a normalization factor.

Finally since $P(x)$ is proportional to the amplitude of the  $H_0$ mode, that is the Maxwellian,
we can write the following approximate steady state  phase-space distribution function,
corresponding to the state with vanishing currents:

\be
f(x,v)= {\cal N} \sqrt{\frac{\tau}{2\pi D}}   \sqrt{\frac{1}{1-\tau F'(x)/\gamma}}  \exp \Bigl( \frac{1}{1-\tau F'(x)/\gamma} \frac{\tau} {D} v^2\Bigl) P(x)
\ee


 \section{Evaluation of  the determinant for large $N$}
 \label{appendixdeterminant}  
 The exact evaluation of the determinant of the matrix $\Gamma$ and of its inverse 
 is a formidable task and is far beyond the authors capabilities. However, it is possible to
  provide an approximate matrix  inversion  by expanding 
to linear order in $\tau/\gamma$ the formulas. 
In order to illustrate the point, we consider
the matrix in the case of $N$ particles in  two spatial dimensions:
$$
\begin{small}
\left(\begin{array}{ccccccc}
[1+\frac{\tau}{\gamma}\sum_{j\neq 1} w_{xx}(\rr_1,\rr_j)]& \sum_{j\neq 1} \frac{\tau}{\gamma}w_{xy}(\rr_1,\rr_j)& -\frac{\tau}{\gamma}w_{xx}(\rr_1,\rr_2) &\dots&-\frac{\tau}{\gamma}w_{xy}(\rr_1,\rr_N) 
\\ \sum_{j\neq 1} \frac{\tau}{\gamma}w_{yx}(\rr_1,\rr_j)   & [1+\frac{\tau}{\gamma}\sum_{j\neq 1} w_{yy}(\rr_1,\rr_j)] & -\frac{\tau}{\gamma}w_{yx}(\rr_1,\rr_2)&\dots     &-\frac{\tau}{\gamma}w_{yy}(\rr_1,\rr_N)
\\ -\frac{\tau}{\gamma}w_{xx}(\rr_2,\rr_1)& -\frac{\tau}{\gamma}w_{yx}(\rr_2,\rr_1) & [1+\frac{\tau}{\gamma}\sum_{j\neq 2}  w_{xx}(\rr_2,\rr_j)] & \dots&-\frac{\tau}{\gamma}w_{xy}(\rr_2,\rr_N) \\
\\\dots&\dots&\dots&\dots&\dots&\\
\\ -\frac{\tau}{\gamma}w_{xy}(\rr_N,\rr_1)& -\frac{\tau}{\gamma}w_{yy}(\rr_N,\rr_1)&\dots & \dots& [1+\frac{\tau}{\gamma}\sum_{j\neq N}  w_{yy}(\rr_N,\rr_j)] 
\end{array}\right)
\end{small}
$$
It is interesting to remark that the off-diagonal elements contain only one term, while the diagonal elements and their
neighbours contain $N$ elements. Thus in the limit of $N\to\infty$ we expect that the matrix becomes effectively diagonal.

$$
\begin{small}
\left(\begin{array}{ccccccc}
[1+\frac{\tau}{\gamma}\sum_{j\neq 1} w_{xx}(\rr_1,\rr_j)]& \sum_{j\neq 1} \frac{\tau}{\gamma}w_{xy}(\rr_1,\rr_j)& 0 &\dots&0
\\ \sum_{j\neq 1} \frac{\tau}{\gamma}w_{yx}(\rr_1,\rr_j)   & [1+\frac{\tau}{\gamma}\sum_{j\neq 1} w_{yy}(\rr_1,\rr_j)] & 0&\dots     &0
\\
\\\dots&\dots&\dots&\dots&\dots\\
 0& 0 & 0&[1+\frac{\tau}{\gamma}\sum_{j\neq N}  w_{xx}(\rr_N,\rr_j)] &  \sum_{j\neq 2} \frac{\tau}{\gamma}w_{yx}(\rr_N,\rr_j) \\
\\ 0&0&0& \sum_{j\neq N} \frac{\tau}{\gamma}w_{yx}(\rr_N,\rr_j) & [1+\frac{\tau}{\gamma}\sum_{j\neq N}  w_{yy}(\rr_N,\rr_j)] 
\end{array}\right)
\end{small}
$$

Its inverse is approximately
$$
\begin{small}
\left(\begin{array}{ccccccc}
[1-\frac{\tau}{\gamma}\sum_{j\neq 1} w_{xx}(\rr_1,\rr_j)]&- \sum_{j\neq 1} \frac{\tau}{\gamma}w_{xy}(\rr_1,\rr_j)& 0 &\dots&0
\\ -\sum_{j\neq 1} \frac{\tau}{\gamma}w_{yx}(\rr_1,\rr_j)   & [1-\frac{\tau}{\gamma}\sum_{j\neq 1} w_{yy}(\rr_1,\rr_j)] & 0&\dots     &0
\\
\\\dots&\dots&\dots&\dots&\dots\\
 0& 0 & 0&[1-\frac{\tau}{\gamma}\sum_{j\neq N}  w_{xx}(\rr_N,\rr_j)] &  -\sum_{j\neq 2} \frac{\tau}{\gamma}w_{yx}(\rr_N,\rr_j) \\
\\ 0&0&0& -\sum_{j\neq N} \frac{\tau}{\gamma}w_{yx}(\rr_N,\rr_j) & [1-\frac{\tau}{\gamma}\sum_{j\neq N}  w_{yy}(\rr_N,\rr_j)] 
\end{array}\right)
\end{small}
$$
The determinant  to order $\tau/\gamma$ is 
\be
 \det \Gamma\approx1+ \frac{\tau}{\gamma} \sum_{i,j,i\neq j} [w_{xx}(\rr_i,\rr_j)+w_{yy}(\rr_i,\rr_j)  ]
\ee



\newpage

\begin{thebibliography}{22}
\providecommand{\url}[1]{\texttt{#1}}
\providecommand{\urlprefix}{URL }
\markboth{Taylor \& Francis and I.T. Consultant}{Molecular Physics}


\bibitem{bechinger2016}
C. Bechinger, R. Di~Leonardo, H. Lowen, C. Reichhardt, G. Volpe and G. Volpe,
  ArXiv preprint arXiv:1602.00081   (2016).

\bibitem{marchetti2013hydrodynamics}
M. Marchetti, J. Joanny, S. Ramaswamy, T. Liverpool, J. Prost, M. Rao and R.A.
  Simha,  Reviews of Modern Physics  \textbf{85} (3), 1143-1189 (2013).

\bibitem{marchetti2015structure}
M.C. Marchetti, Y. Fily, S. Henkes, A. Patch and D. Yllanes,  arXiv preprint
  arXiv:1510.00425   (2015).

\bibitem{poon2013clarkia}
W. Poon,  Proceedings of the International School of Physics Enrico Fermi,
  Course CLXXXIV Physics of Complex Colloids, eds. C. Bechinger, F. Sciortino
  and P. Ziherl, IOS, Amsterdam: SIF, Bologna  pp. 317--386 (2013).

\bibitem{romanczuk2012active}
P. Romanczuk, M. Baer, W. Ebeling, B. Lindner and L. Schimansky-Geier,  The
  European Physical Journal Special Topics  \textbf{202} (1), 1-162 (2012).

\bibitem{elgeti2014physics}
J. Elgeti, R.G. Winkler and G. Gompper,  arXiv preprint arXiv:1412.2692
  (2014).

\bibitem{cates2012diffusive}
M. Cates,  Reports on Progress in Physics  \textbf{75} (4), 042601-042614 (2012).

\bibitem{bialke2014negative}
J. Bialk{\'e}, H. L{\"o}wen and T. Speck,  arXiv preprint arXiv:1412.4601
  (2014).

\bibitem{fily2012athermal}
Y. Fily and M.C. Marchetti,  Physical Review Letters  \textbf{108} (23), 235702-235706
  (2012).

\bibitem{stenhammar2013continuum}
J. Stenhammar, A. Tiribocchi, R.J. Allen, D. Marenduzzo and M.E. Cates,
  Physical Review Letters  \textbf{111} (14), 145702-145706 (2013).

\bibitem{tailleur2008statistical}
J. Tailleur and M. Cates,  Physical Review Letters  \textbf{100} (21), 218103-218106
  (2008).

\bibitem{cates2013active}
M. Cates and J. Tailleur,  EPL (Europhysics Letters)  \textbf{101} (2), 20010-20015
  (2013).

\bibitem{berg2004coli}
H.C. Berg, \emph{E. coli in Motion}, Springer Verlag, New York   (  2004).

\bibitem{schnitzer1993theory}
M.J. Schnitzer,  Physical Review E  \textbf{48} (4), 2553-2568 (1993).

\bibitem{zheng2013nongaussian}
X. Zheng, B. ten Hagen, A. Kaiser, M. Wu, H. Cui, Z. Silber-Li and H.
  L{\"o}wen,  Physical Review E  \textbf{88} (3), 032304-032314 (2013).

\bibitem{maggi2014generalized}
C. Maggi, M. Paoluzzi, N. Pellicciotta, A. Lepore, L. Angelani and R.
  Di~Leonardo,  Physical Review Letters  \textbf{113} (23), 238303-238307 (2014).

\bibitem{koumakis2014delivery}
N. Koumakis, C. Maggi and R. Di~Leonardo,  Soft Matter  \textbf{10} (31), 5695-5701
  (2014).

\bibitem{koumakis2013targeted}
N. Koumakis, A. Lepore, C. Maggi and R. Di~Leonardo,  Nature communications
  \textbf{4}, 1-6 (2013).

\bibitem{maggi2011effective}
L. Angelani, C. Maggi, M. Bernardini, A. Rizzo and R. Di~Leonardo,  Physical
  Review Letters  \textbf{107} (13), 138302-138305 (2011).

\bibitem{farage2015effective}
T. Farage, P. Krinninger and J. Brader,  Physical Review E  \textbf{91} (4),
  042310-042319 (2015).

\bibitem{marconi2015towards}
U.Marini Bettolo Marconi and C. Maggi,  Soft Matter  \textbf{11} (45), 8768-8781 (2015).

\bibitem{marconi2010dynamic}
U. Marini Bettolo Marconi and S. Melchionna,  Journal of Physics: Condensed Matter
  \textbf{22} (36), 364110-364117 (2010).

\bibitem{hanggi1995colored}
P. Hanggi and P. Jung,  Advances in Chemical Physics  \textbf{89}, 239-326 (1995).

\bibitem{cao1993effects}
L. Cao, D.j. Wu and X.l. Luo,  Physical Review A  \textbf{47} (1), 57-70 (1993).

\bibitem{maggi2015multidimensional}
C. Maggi, U.Marini Bettolo Marconi, N. Gnan and R. Di~Leonardo,  Scientific Reports
  \textbf{5}, 1-7 (2015).

\bibitem{marconi2015velocity}
U.Marini Bettolo Marconi, N. Gnan, C. Maggi, M. Paoluzzi and R. Di~Leonardo,  arXiv
  preprint arXiv:1512.04227   (2015).

\bibitem{purcell1977life}
E.M. Purcell,  Am. J. Phys  \textbf{45} (1), 3-11 (1977).

\bibitem{cates2014motility}
M. Cates and J. Tailleur,  Annual Reviews Condensed Matter Physics  \textbf{6}
  (6), 219-244 (2015).

\bibitem{barre2014motility}
J. Barr{\'e}, R. Ch{\'e}trite, M. Muratori and F. Peruani,  arXiv preprint
  arXiv:1403.2364   (2014).

\bibitem{marconi2006nonequilibrium}
U.Marini Bettolo Marconi and P. Tarazona,  The Journal of Chemical Physics  \textbf{124}
  (16), 164901-164911 (2006).

\bibitem{marini2007}
U. Marini-Bettolo-Marconi, P. Tarazona and F. Cecconi,  The Journal of Chemical
  Physics  \textbf{126}, 164904-164916 (2007).



\end{thebibliography}

\end{document}